\documentclass{emulateapj}

\newcommand{\Lya}{Ly$\alpha$ }
\newcommand{\lya}{Ly$\alpha$}

\begin{document}

\title{The evolution of the helium-ionizing background at $z \sim 2$--$3$}
\author{Keri L. Dixon \& Steven R. Furlanetto}

\affil{Department of Physics and Astronomy, University of California, Los Angeles, CA 90095, USA; email: kdixon@astro.ucla.edu}

\begin{abstract} Recent observations suggest that helium became fully ionized around redshift $z \sim 3$. The \ion{He}{2} optical depth derived from the Lyman-$\alpha$ forest decreases substantially from this period to $z \sim 2$; moreover, it fluctuates strongly near $z \sim 3$ and then evolves smoothly at lower redshifts. From these opacities, we compute, using a semi-analytic model, the evolution of the mean photoionization rate and the attenuation length for helium over the redshift range $2.0 \lesssim z \lesssim 3.2$. This model includes an inhomogeneous metagalactic radiation background, which is expected during and after helium reionization.  We find that assuming a uniform background underestimates the required photoionization rate by up to a factor $\sim 2$. When averaged over the (few) available lines of sight, the effective optical depth exhibits a discontinuity near $z \approx 2.8$, but the measurement uncertainties are sizable. This feature translates into a jump in the photoionization rate and, provided the quasar emissivity evolves smoothly, in the effective attenuation length, perhaps signaling the helium reionization era. We then compute the evolution of the effective optical depth for a variety of simple helium reionization models, in which the measured quasar luminosity function and the attenuation length, as well as the evolving \ion{He}{3} fraction, are inputs. A model with reionization ending around redshift $z \approx 2.7$ is most consistent with the data, although the constraints are not strong thanks to the sparseness of the data. \end{abstract}

\keywords{intergalactic medium -- diffuse radiation -- quasars: absorption lines}

\section{ Introduction } \label{sec:intro}

In the standard reionization history of the Universe, the metagalactic ionizing background evolved relatively slowly except for the reionization of hydrogen ($z \gtrsim 6$) and of helium (fully-ionized at $z \sim 3$). During these phase transitions, the intergalactic medium (IGM) became mostly transparent to the relevant ionizing photons, allowing the high-energy radiation field to grow rapidly. Since \ion{He}{2} has an ionization potential of 54.4~eV, bright quasars with hard spectra are required for its ionization. As such, the distribution and intrinsic character of quasars, in addition to the properties of the IGM, determine the radiation background at these high energies.

These quasars are quite rare, implying a strongly fluctuating background even after reionization \citep{Fard98, Bolt06, Meik07, Furl08b}. Direct evidence for these source-induced variations has been seen in the ``transverse proximity effect" of the hardness ratio through comparisons of the \ion{H}{1} and \ion{He}{2} Lyman-$\alpha$ (\lya) forests with surveys for nearby quasars \citep{Jako03, Wors06, Wors07}. These variations are exaggerated by the strong attenuation from the IGM \citep{Haar96, Fauc08, Furl08}. Furthermore, radiative transfer through the clumpy IGM can induce additional fluctuations \citep{Mase05, Titt07}. During reionization, fluctuations in the background are even greater, because some regions receive strong ionizing radiation while others remain singly-ionized with no local illumination.

Recent observations indicate that helium reionization occurs at $z \sim 3$. The strongest evidence comes from far-ultraviolet spectra of the \ion{He}{2} \Lya forest along the lines of sight to bright quasars at $z \sim 3$. These observations of the \ion{He}{2} \Lya transition ($\lambda_{\rm{rest}} = 304$~\AA) are difficult, because bright quasars with sufficient far-UV flux and no intervening Lyman-limit systems are required. To date, six such lines of sight have yielded opacity measurements: PKS 1935-692 \citep{Tytl95, Ande99}), HS 1700+64 \citep{Davi96, Fech06}, HE 2347-4342 \citep{Reim97, Kris01, Smet02, Shul04, Zhen04}, SDSS J2346-0016 \citep{Zhen04a, Zhen08}, Q0302-003 \citep{Jako94, Hoga97, Heap00, Jako03} and HS 1157-3143 \citep{Reim05}. The effective helium optical depth from these studies decreases rapidly at $z \approx 2.8$, then declines slowly to lower redshifts. The opacities at higher redshifts exhibit a patchy structure with alternating high and low absorption, which may indicate an inhomogeneous radiation background. More promising sightlines  have been detected \citep{Syph09}, and the Cosmic Origins Spectrograph on the \textit{Hubble Space Telescope} should add to the current pool of data.

Several indirect methods attempt to probe the impact of helium reionization on the the IGM. One expected effect of helium reionization is an increase in the IGM temperature \citep{Hui97, Gles05, Furl08a, McQu09}. \citet{Scha00} detected a sudden temperature increase at $z \sim 3.3$ by examining the thermal broadening of \ion{H}{1} \Lya forest lines (see also \citealt{Scha99, Theu02a}). Around the same time, the IGM temperature-density relation appears to become nearly isothermal, another indication of recent helium reionization \citep{Scha00, Rico00}. However, not all studies agree \citep{McDo01}, and temperature measurements via the \Lya forest flux power spectrum show no evidence for any sudden change \citep{Zald01, Viel04, McDo06}. Furthermore, this temperature increase should decrease the recombination rate of hydrogen, decreasing the \ion{H}{1} opacity \citep{Theu02}. Three studies with differing methods have measured a narrow dip at $z \sim 3.2$ in the \ion{H}{1} effective optical depth \citep{Bern03, Fauc08a, Dall08}. While initially attributed to helium reionization \citep{Theu02}, more recent studies find that reproducing this feature with helium reionization is extremely difficult \citep{Bolt09a, Bolt09, McQu09}.

The (average) metagalactic ionizing background should also harden as helium is reionized, because the IGM would become increasingly transparent to high-energy photons. Observations of the \ion{He}{2}/\ion{H}{1} ratio are qualitatively consistent with reionization occurring at $z \sim 3$ in at least one line of sight \citep{Heap00}. \citet{Song98, Song05} found a break in the ratio of \ion{C}{4} to \ion{Si}{4} at $z \sim 3$. Modeling of the ionizing background from optically thin and optically thick metal line systems also shows a significant hardening at $z \sim 3$ \citep{Vlad03, Agaf05, Agaf07}, but other data of comparable quality show no evidence for rapid evolution \citep{Kim02, Agui04}.  However, this approach is made more difficult by the large fluctuations in the \ion{He}{2}/\ion{H}{1} ratio even after helium reionization is complete \citep{Shul04}.

In this paper, we focus on interpreting the \ion{He}{2} \Lya forest and the significance of the jump in the opacity at $z \approx 2.8$. After averaging the effective optical depth over all sightlines, we calculate the expected photoionization rate given some simple assumptions. In particular, we investigate the impact of a fluctuating radiation background, comparing it to the common uniform assumption. We interpret our results in terms of an evolving attenuation length for helium-ionizing photons $R_0$ as well as state-of-the-art models of inhomogeneous reionization.
 
We use a semi-analytic model, outlined in \S\ref{sec:method}, to infer the helium photoionization rate from the \ion{He}{2} \Lya forest. The helium opacity measurements in the redshift range $2.0 \lesssim z \lesssim 3.2$, which serve as the foundation for our calculations, are compiled from the literature in \S\ref{sec:data}. First, we assume a post-reionization universe over the entire redshift span. In this regime, we find the photoionization rate and attenuation length in \S\ref{sec:results}, given the average measured opacity. Motivated by these results, we examine some fiducial reionization histories in \S\ref{sec:toy}. We conclude in \S\ref{sec:disc}.

In our calculations, we assume a cosmology with $\Omega_m = 0.26, \Omega_{\Lambda} = 0.74, \Omega_b = 0.044, H_0 = h$(100~km s$^{-1}$ Mpc$^{-1})$ (with $h$ = 0.74), $n = 0.95$, and $\sigma_8 = 0.8$, consistent with the most recent measurements \citep{Dunk09, Koma09}. 

\section{ Method } \label{sec:method}

The helium \Lya forest observed in the spectra of quasars originates from singly-ionized helium gas in the IGM. Quantitative measurement of this absorption is typically quoted as the transmitted flux ratio $F$, defined as the ratio of observed and intrinsic fluxes, or the related effective optical depth
	\begin{equation} \label{eq:taueff} \tau_{\rm{eff}} \equiv -\ln\langle F \rangle. \end{equation}
We have $F = e^{-\tau_{\rm{eff}}} = \langle e^{-\tau} \rangle > e^{-\langle \tau \rangle}$, thus $\tau_{\rm{eff}} < \langle \tau \rangle$. From the current opacity measurements, we wish to infer the \ion{He}{2} photoionization rate. This connection depends on the details of the IGM, including the temperature, density distribution, and ionized helium fraction. 

\subsection{ Fluctuating Gunn-Peterson approximation } \label{sec:FGPA}

The Gunn-Peterson (\citeyear{Gunn65}) optical depth for \ion{He}{2} \Lya photons is
	\begin{equation} \label{eq:GP} \tau_{GP} = \frac{\pi e^2}{m_ec}f_{\alpha}\lambda_{\alpha}H^{-1}(z)n_{\rm{HeII}}. \end{equation}
Here, the oscillator strength $f_{\alpha} = 0.416$, $\lambda_{\alpha} = 304~\mbox{\AA}$, and $n_{\rm{HeII}}$ is the density of singly-ionized helium in the IGM. For simplicity, we approximate the Hubble constant as $H(z) \approx H_0\Omega_m^{1/2}(1 + z)^{3/2}$. Since the \Lya forest probes the low-density, ionized IGM, most of the hydrogen (mass fraction $X = 0.76$) and helium ($Y = 0.24$) are in the form of \ion{H}{2} and \ion{He}{3}, respectively, after reionization. Under these assumptions, photoionization equilibrium requires
	\begin{equation} \label{eq:balance} \Gamma  n_{\rm{HeII}} = n_{\rm{He}}n_e\alpha_A, \end{equation}
where $\Gamma$ is the \ion{He}{2} photoionization rate and the case-A recombination coefficient is $\alpha_A = 3.54\times10^{-12}(T/10^4~\rm{K})^{-0.7}$~cm$^3$~s$^{-1}$ according to \citet{Stor95}. For a clumpy universe, most photons emitted by recombinations are produced and subsequently reabsorbed in dense, mostly neutral systems. These ionizing photons, therefore, cannot escape to the low density regions of interest for the forest, so we use case-A \citep{Mira03}.
	
The \Lya forest, and therefore the optical depth, trace the local overdensity $\Delta$ of the IGM, where $\Delta \equiv \rho/\bar{\rho}$ and $\bar{\rho}$ is the mean mass density. Since $n_e \propto n_{\rm{He}}$ in a highly ionized IGM, equation~(\ref{eq:balance}) implies that $n_{\rm{HeII}} \propto n^2_{\rm{He}} \propto \Delta^2$. The optical depth is proportional to $n_{\rm{HeII}}$ (see eq.~\ref{eq:GP}), which introduces a $\Delta^2$ factor. Additionally, the temperature of the IGM, which affects the recombination rate, is typically described by a power law of the form $T = T_0\Delta^{\gamma-1}$ \citep{Hui97}, where $T_0$ and $\gamma$ are taken as constants.\footnote{ In actuality, during and after helium reionization both $T_0$ and $\gamma$ likely become redshift- and density-dependent.  We will consider such effects in \S\ref{sec:semi} and \S\ref{sec:T}.} Including the above equations and cosmological factors,
	\begin{eqnarray} \tau_{\rm{GP}} & \simeq & \frac{19}{\Gamma_{-14}}\left(\frac{T_0}{10^4~\rm{K}}\right)^{-0.7}\left(\frac{\Omega_bh^2}{0.0241}\right)^2\left(\frac{\Omega_mh^2}{0.142}\right)^{-1/2} \nonumber \\
	& & \times \left(\frac{1+z}{4}\right)^{9/2}\Delta^{2-0.7(\gamma-1)}, 
	\label{eq:tau}
	\end{eqnarray}
where $\Gamma = 10^{-14}\Gamma_{-14}$~s$^{-1}$.

The fluctuating Gunn-Peterson approximation (FGPA) (e.g., \citealt{Wein99}), equation~(\ref{eq:tau}), relates the effective optical depth, or the continuum normalized flux, to the local overdensity $\Delta$ and the photoionization rate $\Gamma_{-14}$.  This approximation shows the relationship between the opacity and the IGM, but it ignores the effects of peculiar velocities and thermal broadening on the \Lya lines. In practice, when comparing to observations an overall proportionality constant, $\kappa$, is introduced to the right hand side of equation~(\ref{eq:tau}) to compensate for these factors, as described in \S\ref{sec:UVbg}. This normalization also incorporates the uncertainties in $T_0, \Omega_b, \Omega_m$, and $h$.

\subsection{ A semi-analytic model for \Lya absorption } \label{sec:semi}

To fully describe helium reionization and compute the detailed features of the \Lya forest, complex hydrodynamical simulations of the IGM, including radiative transfer effects and an inhomogeneous background, are required. Recent simulations \citep{Soka02, Pasc07, McQu09} have made great advances to incorporate the relevant physics and to increase in scale. The simulations remain computationally intensive, and they cannot simultaneously resolve the $\sim 100$~Mpc scales required to adequately study inhomogeneous helium reionization and the much smaller scales required to self-consistently study the \Lya forest, necessitating some sort of semi-analytic prescription to describe baryonic matter on small scales.  On the other hand, the semi-analytic approach taken here, including fluctuations in the ionizing background, should broadly reproduce the observed optical depth, especially considering the uncertainties in the measurements and the limited availability of suitable quasar lines of sight. To outline, the model has four basic inputs: the IGM density distribution $p(\Delta)$, the temperature-density relation $T(\Delta)$, the radiation background distribution $f(J)$, and the mean helium ionized fraction $\bar{x}_{\rm{HeIII}}$.

\citet{Mira00} suggest the volume-weighted density distribution function
	\begin{equation} p(\Delta) = A\Delta^{-\beta}\exp\left[-\frac{\left(\Delta^{-2/3} - C_0\right)^2}{2\left(2\delta_0/3\right)^2}\right], \end{equation}
where $\delta_0 = 7.61/(1 + z)$ and $\beta$ (for a few redshifts) are given in Table~1 of their paper. Intermediate $\beta$ values were found using polynomial interpolation. The remaining constants, $A$ and $C_0$, were calculated by normalizing the total volume and mass to unity at each redshift. The distribution matches cosmological simulations reasonably well for the redshifts of interest,\footnote{More recent simulations by \citet{Pawl09} and \citet{Bolt09b} basically agree with the above $p(\Delta)$ for low densities, which the \Lya forest primarily probes.} 
i.e. $z = 2 - 4$. Although this form does not incorporate all the physics of reionization and was not generated with the current cosmological parameter values, the overall behavior should be sufficient for the purposes of our model.

A current topic of discussion is the thermal evolution of the IGM during and after helium reionization. The temperature-density relation should vary as a function of redshift and density; notably, helium reionization should increase the overall temperature by a factor of a few \citep{Hui97, Gles05, Furl08a, McQu09}, which may have been observed \citep{Scha00,Rico00}. As mentioned in \S\ref{sec:FGPA}, the temperature is assumed to follow a power law $T = T_0\Delta^{\gamma-1}$ in the FGPA.  Generally, $T_0 \sim 1-2\times10^4$~K and $1 \leq \gamma \leq 1.6$ should broadly describe the post-reionization IGM (e.g., \citealt{Hui97}), but the exact values are a matter of debate. Unless otherwise noted, we use $T_0 = 2\times10^4$~K and $\gamma = 1$ throughout our calculations. The isothermal assumption also suppresses the temperature, and therefore density, dependence of the recombination coefficient $\alpha_A$.

A uniform radiation background has been a common assumption in previous studies, but the sources (quasars) for these photons are rare and bright. Therefore, random variations in the quasar distribution create substantial variations in the high-frequency radiation background \citep{Meik07, Furl08, McQu09}. Furthermore, the $1/r^2$ intensity profiles of these sources induce strong small-scale fluctuations \citep{Furl09a}, which may in turn significantly affect the overall optical depth of the \Lya forest. For the probability distribution $f(J)$ of the angle-averaged specific intensity of the radiation background $J$, we follow the model presented in \citet{Furl08b}.  In the post-reionization limit, the probability distribution can be computed exactly for a given quasar luminosity function and attenuation length, assuming that the sources are randomly distributed (following Poisson statistics).  This distribution can be derived either via Markov's method \citep{Zuo92} or via the method of characteristic functions \citep{Meik03}.

During reionization, the local \ion{He}{3} bubble radius, i.e. the horizon within which ionizing sources are visible, varies across the IGM and so becomes another important parameter. Due to the rarity of sources, with typically only a few visible per \ion{He}{3} region, a Monte Carlo treatment best serves this regime \citep{Furl08b}.  For a given bubble of a specified size, we randomly choose the number of quasars inside each bubble according to a Poisson distribution. Each quasar is then randomly assigned a location within the bubble as well as a luminosity (via the measured luminosity function). Next, we sum the specific intensity from each quasar.  After $10^6$ Monte Carlo trials, this procedure provides $f(J)$ for a given bubble size; the solution converges to the post-reionization scenario for an infinite bubble radius.  The final ingredient is the size distribution of discrete ionized bubbles, which is found using the excursion set approach of \citet{Furl08} (based on the hydrogen reionization equivalent from \citealt{Furl04}).  After integrating over all possible bubble sizes, $f(J)$ depends on the parameters $z, \, R_0,$ and $\bar{x}_{\rm{HeIII}}$, in addition to the specified luminosity function.

In the following calculations, we scale $J$ (which is evaluated at a single frequency) to $\Gamma$ (which integrates over all frequencies) simply using the ratio of the mean photoionization rate to the mean radiation background.  This is not strictly correct, because higher frequency photons have larger attenuation lengths and so more uniform backgrounds; however, it is a reasonable prescription because the ionization cross section falls rapidly with photon frequency. However, it does mean that we ignore the large range in spectral indices of the ionizing sources (see below), which modulate the shape of the local ionizing background and lead to an additional source of fluctuations in $\Gamma$ relative to $J$ that we do not model.
The largest problem occurs during helium reionization, when the highest energy photons can travel \emph{between} \ion{He}{3} regions (a process we ignore); however, they have small ionization cross sections and so do not significantly change our results, except very near the end of that process \citep{Furl08b}.

\begin{figure*}
    \plottwo {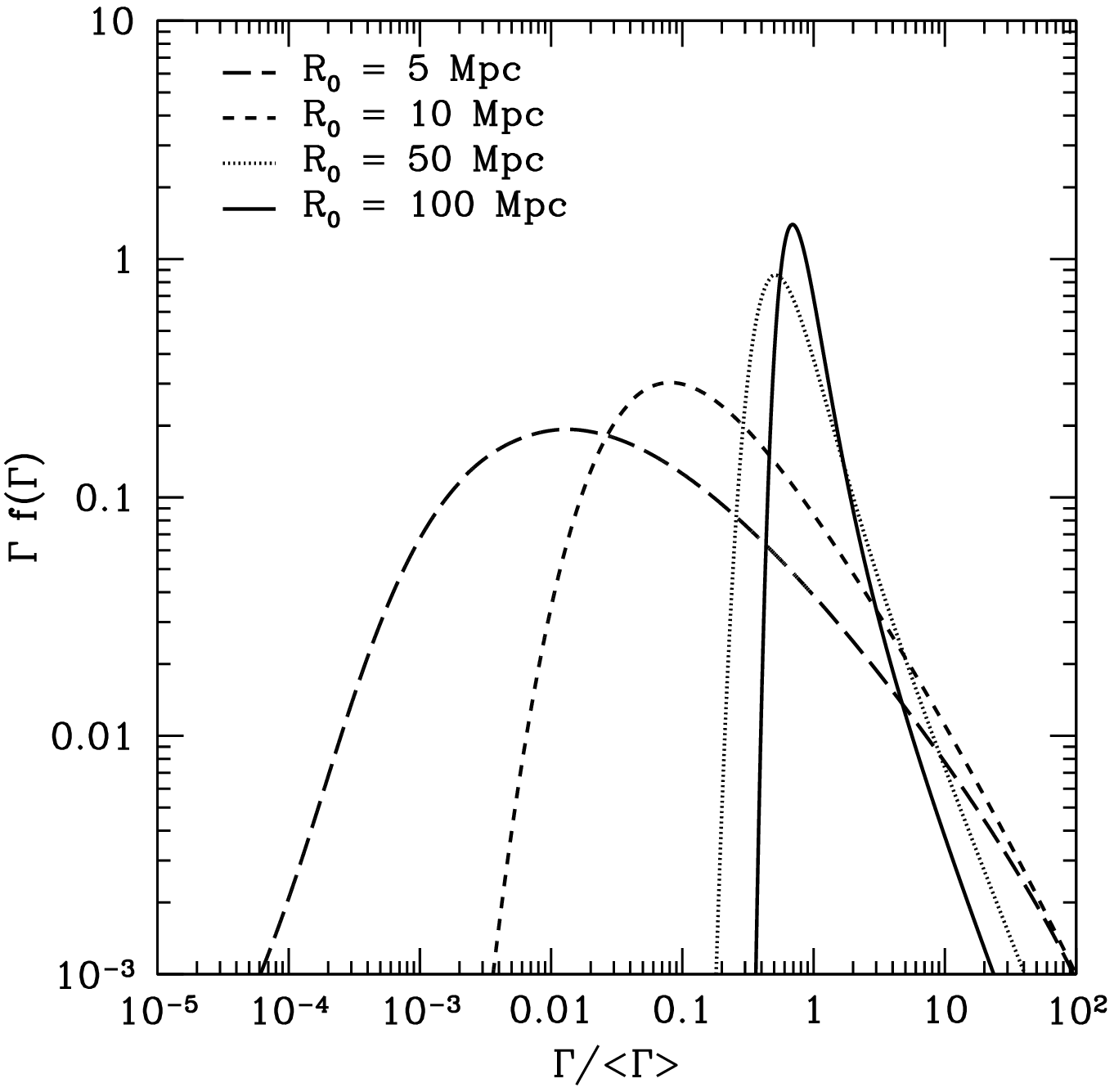} {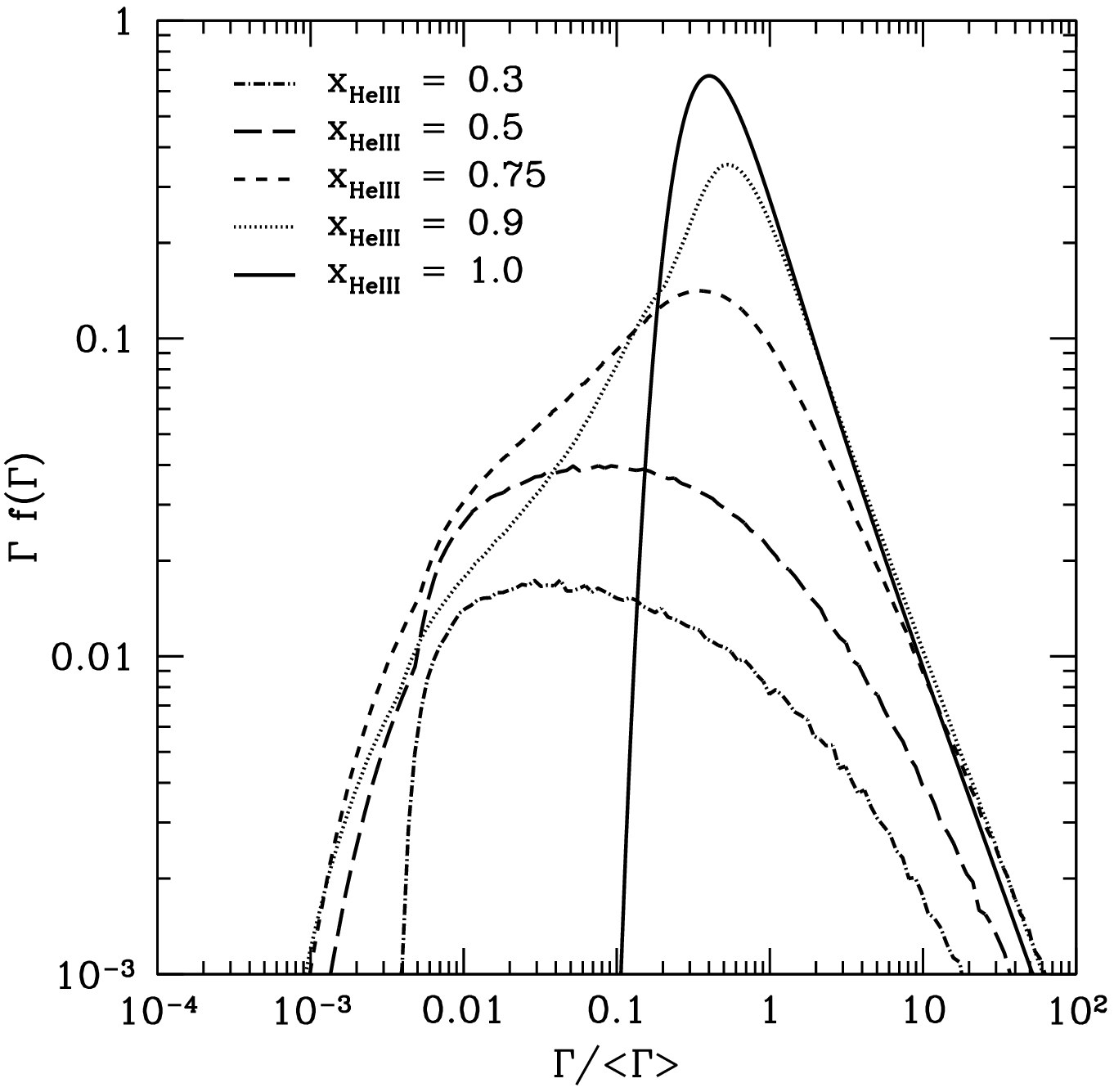}
  \caption{ Distribution of the photoionization rate $\Gamma$ relative to its mean value in a fully-ionized IGM, $\langle \Gamma \rangle $, at redshift $z = 2.5$. \textit{Left panel:} The curves assume $R_0 = 5, 10, 50,$ and 100~Mpc, from widest to narrowest, in a post-reionization universe. \textit{Right panel:} The curves take $\bar{x}_{\rm{HeIII}} = 0.3, 0.5, 0.75, 0.9,$ and 1.0, from lowest to highest at peak, for $R_0 = 35$~Mpc.}
  \label{fig:f(j)}
\end{figure*}

For the majority of the paper, we consider the IGM to be fully-ionized, i.e. $\bar{x}_{\rm{HeIII}} = 1.0$. In this post-reionization regime, $R_0$, the attenuation length of the ionizing photons, determines the shape of $f(\Gamma)$, given the redshift and other model assumptions. We show several example distributions in the left panel Figure~\ref{fig:f(j)}. From widest to narrowest, the curves correspond to $R_0 = 5, 10, 50,$ and 100~Mpc ($z = 2.5$). Smaller attenuation lengths yield a greater spread in $\Gamma$. Qualitatively, more sources contribute to ionizing a given patch of the IGM for higher $R_0$, making the peak photoionization rate more likely, i.e. the curve is narrower. A uniform background corresponds to $R_0 \rightarrow \infty$. Although low $\Gamma$ values are more likely for low $R_0$, the high-$\Gamma$ tail is nearly independent of $R_0$. This is because a large $\Gamma$ occurs within the ``proximity zone" of a single quasar, making it relatively independent of contributions from much larger scales (unless $R_0$ is much smaller than the proximity zone itself).
During reionization, as in \S\ref{sec:toy}, the ionized fraction \textit{and} the mean free path affect the distribution function, as shown in the right panel of Figure~\ref{fig:f(j)}. Here, we take $z = 2.5$, $R_0 = 35$~Mpc, and $\bar{x}_{\rm{HeIII}} = 0.3, 0.5, 0.75, 0.9,$and 1.0 (from lowest to highest at the peak). A broader distribution of photoionization rates is expected, because the large spread in \ion{He}{3} bubble sizes restricts the source horizon inhomogeneously across the Universe.

To estimate $\tau_{\rm{eff}}$ (or $F$), we integrate over all densities and photoionization rates:
	\begin{equation} \label{eq:F} F = e^{-\tau_{\rm{eff}}} = \int_0^{\infty} d\Gamma f(\Gamma) \int_0^{\infty} d\Delta e^{-\tau(\Delta | \Gamma)}p(\Delta).  \end{equation}
This integral is valid if $\Gamma$ and $\Delta$ are uncorrelated. Since relatively rare quasars ionize \ion{He}{2}, random fluctuations in the number of sources (as opposed to their spatial clustering) dominate the ionization morphology \citep{McQu09}, justifying our assumption of an uncorrelated IGM density and the photoionization rate.

\subsection{ The UV background from quasars } \label{sec:UVbg}

Since quasars ionize the \ion{He}{2} in the IGM, the UV metagalactic background can, in principle, be calculated directly from distribution and intrinsic properties of quasars. Currently, these details, i.e. the quasar luminosity function and attenuation length, are uncertain. The following method for estimating $\Gamma$ is used as a reference for the semi-analytic model. The \ion{He}{2} ionization rate is
	\begin{equation} \label{eq:Gamma} \Gamma = 4\pi\int_{\nu_{\rm{HeII}}}^{\infty} \frac{J_{\nu}}{h\nu}\sigma_{\nu}d\nu, 	\end{equation}
where $\sigma_{\nu} = 1.91\times10^{-18}(\nu/\nu_{\rm{HeII}})^{-3}$~cm$^2$ is the photoionization cross section for \ion{He}{2} and $\nu_{\rm{HeII}}$ is the photon frequency needed to fully ionize helium. For the radiation background at frequency $\nu$, $J_{\nu}$, we assume a simplified form, the absorption limited case in \citet{Meik03}:
	\begin{equation} \label{eq:J} J_{\nu} = \frac{1}{4\pi}\epsilon_{\nu}(z)R_0(z), \end{equation}
where $\epsilon_{\nu}$ is the quasar emissivity and $R_0$ is the attenuation length. 

\begin{figure}
   \plotone {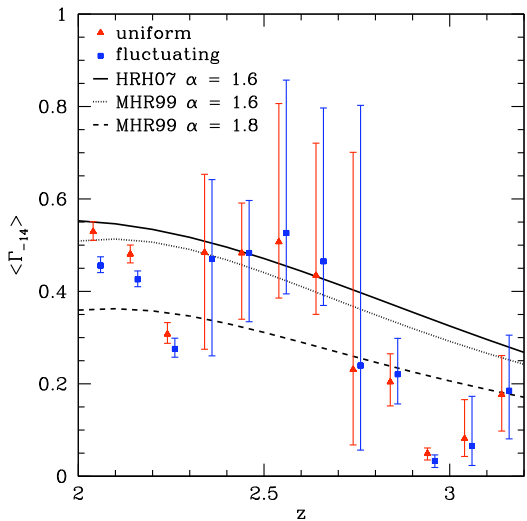} 
    \caption{Evolution of the mean photoionization rate (in units of $10^{-14}$~s$^{-1}$) with redshift. The solid curve represents the inferred ionization rate from the HRH07 quasar luminosity function with $\alpha = 1.6$. The dotted (dashed) curve follows \citet{Mada99} with extreme-UV spectral index $\alpha = 1.6~(1.8)$. All take the attenuation length from eq.~(\ref{eq:Ro}).  The points (with a slight redshift offset) show the reconstructed photoionization rate given the measured effective optical depth, assuming a uniform (triangles) and fluctuating (squares) radiation background. The results are fixed to the solid curve at $z = 2.45$.}
   \label{fig:Gamma}
\end{figure}

We begin with the $B$-band emissivity $\epsilon_{B}$, derived from the quasar luminosity function in \citet[hereafter HRH07]{Hopk07}. To convert this to the extreme-UV (EUV) frequencies of interest, we follow a broken power-law spectral energy distribution \citep{Mada99}:
	\begin{equation} L(\nu) \propto   \left\{ \begin{array}{ll}
	\nu^{-0.3}		& 	~~~2500 < \lambda < 4600~\mbox{\AA}\\
	\nu^{-0.8}		& 	~~~1050 < \lambda < 2500~\mbox{\AA}\\
	\nu^{-\alpha}	& 	~~~\lambda < 1050~\mbox{\AA}. \end{array} \right. \end{equation}
The EUV spectral index $\alpha$ is a source of debate and is not the same for all quasars. \citet{Telf02} find a wide range of values for individual quasars, e.g. $\alpha$ = -0.56 for HE~2347-4342 and 5.29 for TON~34. Most quasars lie closer to the mean, but HE~2347-4342 is a \ion{He}{2} \Lya line of sight in \S\ref{sec:data}. Unless otherwise noted, we use the mean value $\langle \alpha \rangle \approx 1.6$ from \citet{Telf02}, ignoring the variations in $\alpha$. \citet{Zhen97} found $\langle \alpha \rangle \approx 1.8$, which serves as a comparison. The uncertainty in the spectral index $\alpha$ affects the amplitude, not the shape, of the $\Gamma$ curve derived from the QLF (see Fig.~\ref{fig:Gamma}). Since our semi-analytic calculations are normalized to a single point on the $\Gamma$ curve, this uncertainty translates into an amplitude shift in our results, i.e. changes $\kappa$.

The next ingredient for $J_{\nu}$ is the attenuation length of helium-ionizing photons $R_0$. As described above, $R_0$ depends on the photon frequency; for example, high-energy photons can propagate larger distances.  For simplicity, we use a single frequency-averaged attenuation length and focus only on the redshift evolution (in any case, the absolute amplitude can be subsumed into our normalization factor $\kappa$ below).  For concreteness, we apply the $comoving$ form found in \citet{Bolt06}:
	\begin{equation} \label{eq:Ro} R_0 = 30\left(\frac{1 + z}{4}\right)^{-3}~\rm{Mpc}, \end{equation}
which assumes the number of Lyman limit systems per unit redshift is proportional to $(1 + z)^{1.5}$ \citep{Stor94} and uses the normalization based on the model of \citet{Mira00}. An alternate approach, probably more appropriate during reionization itself, is to estimate the attenuation length around individual quasars, as in \citet{Furl08}. This method gives a similar value at $z=3$ but with a \emph{slower} redshift evolution, which would only strengthen our conclusions. This empirical equation for the mean free path (with the above quasar emissivity) provides the photoionization rate $\Gamma$, shown in Figure~\ref{fig:Gamma}. The photoionization rate from the \citet{Mada99} QLF (with $\alpha = 1.6$ and 1.8) is also included in the figure to illustrate the effect of uncertainties in the quasar properties. The spectral index has a greater effect on the amplitude than the differing QLFs.

As noted in \S\ref{sec:FGPA}, comparing our FGPA treatment to the real \Lya forest data requires an uncertain correction to the $\tau-\Delta$ relation in equation~\ref{eq:tau}. For this purpose, we assume the above emissivity and attenuation length to be accurate. Then, the semi-analytic model is adjusted so that the predicted $\tau_{\rm{eff}}$ matches the measured value at a particular redshift. To do so, we insert a prefactor, $\kappa$, to the right hand side of equation~(\ref{eq:tau}). This factor compensates for line blending and other detailed physics ignored by the FGPA, but it also includes any uncertainties in the underlying cosmological or IGM parameters (such as $T_0$). A suitable normalization redshift should be after reionization and have data from more than one line of sight (see the right panel of Fig.~\ref{fig:expt}). Throughout this work, we take $z = 2.45$ as our fiducial point. 

If $\kappa$ is a constant with redshift, our choice of reference point mainly affects the overall amplitude of the photoionization rate or attenuation length, not the redshift evolution. However, since $\kappa$ depends on the IGM properties, it may change with redshift, density, and/or temperature. For example, an increase in temperature broadens the widths of the absorption lines in the \Lya forest, decreasing the importance of saturation but increasing the likelihood of line blending.  Reionization, a drastic change to the IGM, should also affect $\kappa$. Here, we take $\kappa$ to be independent of $z$. The precise value lies somewhere between 0.1 and 0.5, depending on the specific model. \citet{Furl09b} find $\kappa = 0.3$ for the post-reionization hydrogen \Lya effective optical depth, which is similar to our values. In any case, we emphasize that our method \emph{cannot} be used to estimate the absolute value of the ionizing background -- for which detailed simulations are necessary -- but we hope that it can address the redshift evolution of $\Gamma$.

\section{ Evolution of the \ion{He}{2} effective optical depth } \label{sec:data}

\begin{figure*}
    \plottwo {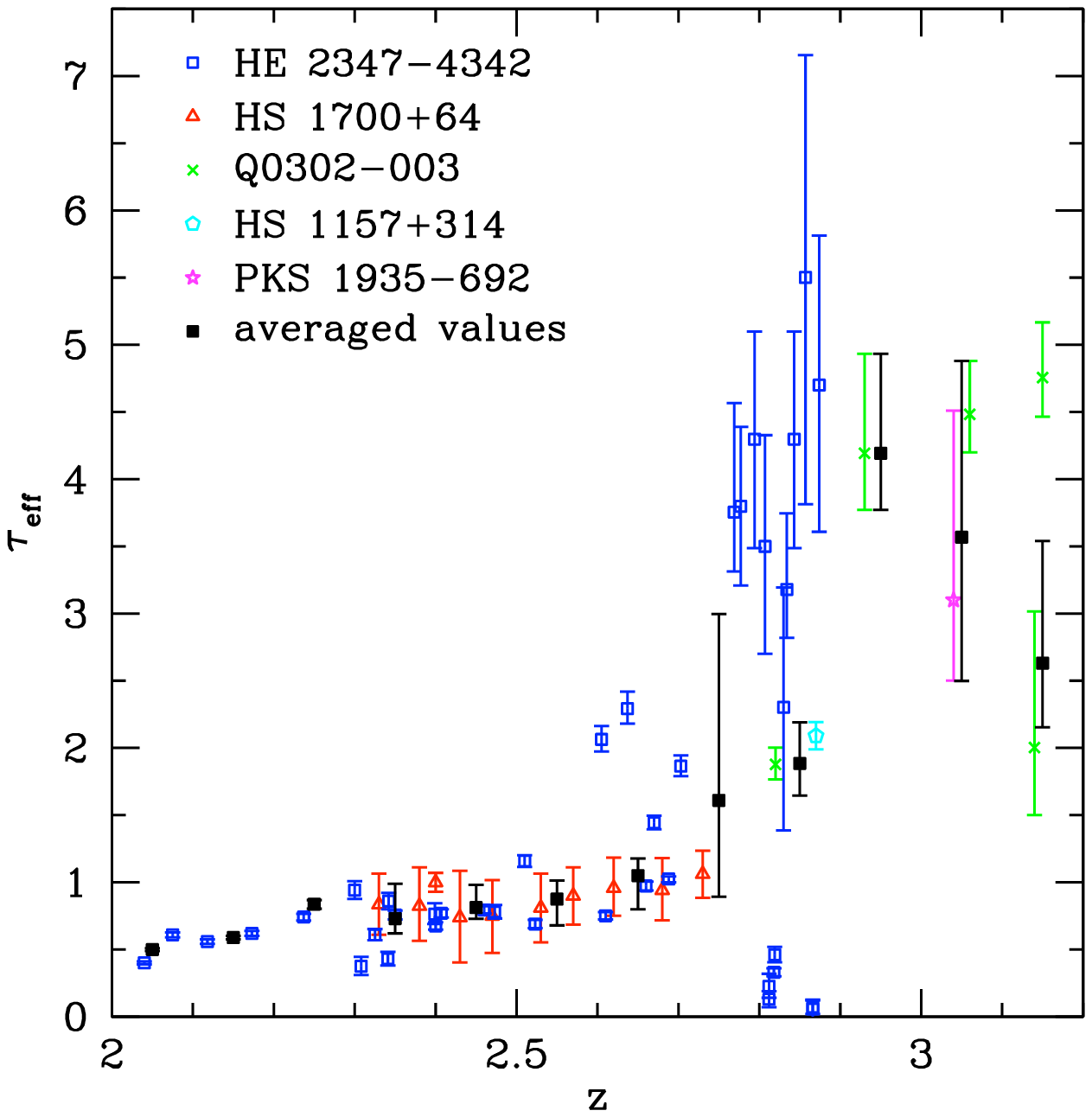} {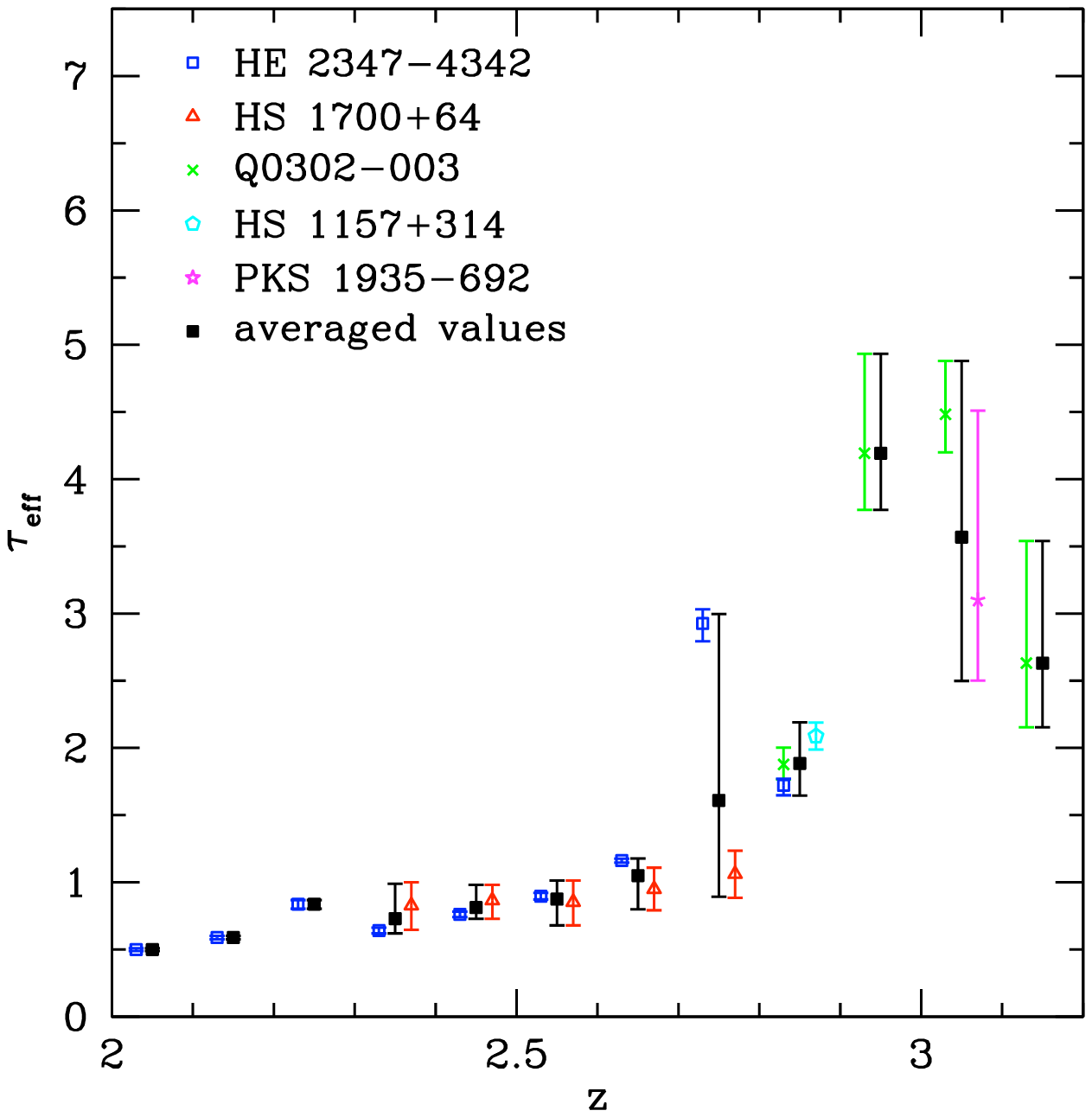}
  \caption{Evolution of the \ion{He}{2} effective optical depth based on the observations of the  \Lya forest for five quasar spectra: HE 2347-4342, HS 1700+64, Q0302-003, HS 1157+314, and PKS 1935-692. The squares are the opacity measurements averaged over redshift bins $\Delta z = 0.1$ with suggestive uncertainties. \textit{Left panel:} Data and uncertainties as quoted in the literature are plotted. \textit{Right panel:} The average values for each line of sight are displayed, elucidating the origin of the uncertainties in the average opacities used throughout the paper. The small redshift offsets within each bin are for illustrative purposes only. }
  \label{fig:expt}
\end{figure*}

Measurements of the \ion{He}{2} effective optical depth are challenging. Suitable lines of sight require a bright quasar with sufficient far-UV flux and no intervening Lyman-limit systems. Currently, only five quasar spectra have provided \ion{He}{2} opacity measurements appropriate for our analysis, displayed in the left panel of Figure~\ref{fig:expt} with the averaged values. \citet{Zhen04a} measure a lower limit on the optical depth at $z \sim 3.5$ from a sixth sightline, SDSS J2346-0016; we do not include any lower limits in any subsequent calculations, but we reference this measurement in \S\ref{sec:toy}. Due to the limited scope of the data, the observed opacities may not be representative of the IGM as a whole.

\textit{HE 2347-4342:} This quasar ($z_{\rm{em}} = 2.885$) is especially bright; therefore, this line of sight has been extensively analyzed. \citet{Zhen04} completed the most comprehensive investigation, covering the redshift range $2.0 < z < 2.9$, including \Lya and Ly$\beta$. \citet{Zhen04} and \citet{Shul04} ($2.0 < z < 2.9$ and \Lya only) utilized high-resolution spectra from the Far Ultraviolet Spectroscopic Explorer (FUSE; $R \sim 20,000$) and the Very Large Telescope (VLT; $R \sim 45,000$). An older and lower resolution study, \citet{Kris01}, covered $2.3 < z < 2.7$ with FUSE. Below redshift $z = 2.7$, the effective helium optical depth evolves smoothly. At higher redshifts, the opacity exhibits a patchy structure with very low and very high absorption, often described respectively as voids and filaments in the literature.

\textit{HS 1700+64:} The \Lya forest of this quasar ($z_{\rm{em}} = 2.72$) has been resolved with FUSE over the redshift range $2.29 \lesssim z \lesssim 2.75$ \citep{Fech06}. An older study using the Hopkins Ultraviolet Telescope (HUT) is consistent with the newer, higher resolution results \citep{Davi96}. The helium opacity evolves smoothly and exhibits no indication of reionization. 

\textit{Q0302-003:} The spectrum of this quasar ($z_{\rm{em}} = 3.286$) was observed with the Space Telescope Imaging Spectrograph (STIS) aboard the Hubble Space Telescope (HST) at 1.8~\AA~resolution \citep{Heap00}. The effective optical depth generally increases with increasing redshift over $2.77 \lesssim z \lesssim 3.22$, excluding a void at $z \sim 3.05$ due to a nearby ionizing source \citep{Bajt88, Zhen95, Giro95}. The data were averaged over redshift bins of $\Delta z \simeq 0.1$. \citet{Hoga97} presented an analysis using the Goddard High Resolution Spectrograph (GHRS) which generally agrees with the later study but quoted a noticeably lower \ion{He}{2} optical depth near $z \sim  3.15$.

\textit{HS 1157+314:} \citet{Reim05} obtained low resolution HST/STIS spectra of the \ion{He}{2} \Lya forest toward this quasar ($z_{\rm{em}} \sim 3$).  Over the redshift range ($2.75 \leq z \leq 2.97$) of the study a patchy structure, similar to HE 2347-4342, is present. The given optical depth was averaged over a redshift bin of $\Delta z \simeq 0.1$.

\textit{PKS 1935-692:} The HST/STIS spectrum for this quasar ($z_{\rm{em}} = 3.18$) was analyzed by \citet{Ande99}. Only one optical depth was quoted, but the spectrum exhibited the usual fluctuations.   

To merge these data sets, we initially binned them in redshift intervals of 0.1, starting with $z = 2.0$. Each bin was assigned the median redshift value, e.g. $z = 2.35$ for $2.3 \leq z < 2.4$. To objectively combine the data, the transmission flux ratios for each data set were averaged (weighted by redshift coverage) in the redshift bins. Then, the values for each quasar were averaged, since multiple data sets may cover the same line of sight. Finally, if more than one line of sight contributes to a bin, the fluxes are averaged once again. The process translates the left panel to the right panel of Figure~\ref{fig:expt}. The uncertainties along each line of sight in the right panel are simply the errors from each separate point in the literature, added in quadrature, without regard to systematic errors; our averaged values then take errors equal to the range spanned by these separate lines of sight. For the remainder of the paper, the error bars result from assuming $F \pm dF$ for each redshift bin, where $F$ are the squares and $\pm dF$ are the upper/lower error bars on the squares in the figure. The small number of well-studied lines of sight limits the amount of truly quantitative statements that can be made.

\section{ Results } \label{sec:results}

\subsection{ Mean \ion{He}{2} photionization rate } \label{sec:Gamma}

We now apply our semi-analytic model to the observed \ion{He}{2} opacity found in the right panel of Figure~\ref{fig:expt}. For each redshift bin, the photoionization rate is calculated by iteratively solving equation~(\ref{eq:F}), i.e. varying $\langle \Gamma \rangle$ (the mean photoionization rate for a fully-ionized IGM) until $F$ matches the measurements. As noted in \S\ref{sec:semi}, a mean free path (eq.~\ref{eq:Ro}) is needed to determine $f(\Gamma)$ for the fluctuating background, but the uniform case has no such requirement except insofar as it affects $\langle \Gamma \rangle$. 

\begin{figure}
   \plotone{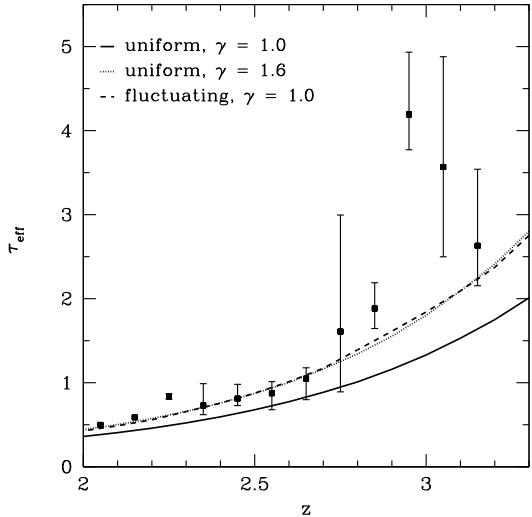}
   \caption{The \ion{He}{2} effective optical depth, assuming the HRH07 QLF and eq.~(\ref{eq:Ro}),  as a function of redshift. The normalization for all curves is constant, $\kappa = 0.291$. The solid (dotted) curve is based on a uniform radiation background, with $\gamma = 1.0~(1.6)$. The dashed curve represents a fluctuating background with an isothermal temperature-density relation. The average measured opacities are shown for reference.}
   \label{fig:kappa}
\end{figure}

First, to provide some intuition, we fix $\kappa$ to be the same for both cases (here, $\kappa = 0.291$) and plot the resulting effective optical depth in Figure~\ref{fig:kappa}, given the emissivity and mean free path from \S\ref{sec:UVbg}. The figure shows the redshift evolution of this optical depth for uniform (solid) and fluctuating (dashed) radiation backgrounds. Interestingly, $\tau_{\rm eff}$ is significantly smaller for a uniform background, especially at higher redshifts (probably due to the shrinking attenuation length assumed in the fluctuating model). This is because most points in the IGM have $\Gamma < \langle \Gamma \rangle$ for the fluctuating background, so most of the IGM has a \emph{higher} opacity; the ``proximity zones" around each quasar are not sufficient to compensate for this effect.  The measured opacities are included in the figure for reference, showing that the shapes of both (normalized) models are consistent with observations for $z < 2.8$ but deviate significantly at higher redshifts.

In practice, $\tau_{\rm eff}$ is the measured quantity, from which we try to infer $\Gamma$.  We find that, for the fiducial attenuation lengths, including a realistic fluctuating background increases the required $\langle \Gamma \rangle$ by about a factor of two -- a nontrivial effect that is important for reconciling quasar observations with the forest.  The magnitude of the required adjustment is comparable to that found by \citet{Bolt06}, who used numerical simulations.
However, as we have emphasized above, we cannot use our model to estimate the absolute value of the ionizing background because the FGPA does not fully describe the \Lya forest; instead we need a renormalization factor $\kappa$.  For the remainder of this paper, we therefore fix the photoionization rate at the $z = 2.45$ HRH07 value (together with our fiducial $R_0$), as described in \S\ref{sec:UVbg}. This procedure gives $\kappa = 0.457$ and 0.291 for the uniform and fluctuating background, respectively.  We strongly caution the reader that the remainder of our quoted results will therefore mask the overall amplitude disparity between these two cases, and we will focus on the redshift evolution of $\Gamma$ instead.

Figure~\ref{fig:Gamma} displays the mean photoionization rate for a uniform (triangles) and fluctuating (squares) UV background (normalized to the fiducial point). The curves represent the photoionization rate inferred from various quasar luminosity functions, as described in \S\ref{sec:UVbg}. Since the $z = 2.45$ reference point was chosen arbitrarily, the overall amplitude should not be considered reliable, which is emphasized by the spread in the QLF curves. The difference between the fluctuating and uniform UV background is small and certainly within the uncertainties. The effect of including fluctuations on the photoionization rate is not straightforward. Generally, at lower $z$, the fluctuating $\Gamma$ is smaller than the uniform result, and the opposite is true for higher redshift. The redshift of the crossover between the two behaviors depends on the amplitude of the measured transmission ratio; a higher $F$ decreases this crossover redshift.

We therefore attribute this effect to changing the characteristic overdensity of regions with high transmission, which is larger at lower redshifts because of the Universe's expansion.  In this regime, where the underlying density field itself has a relatively broad distribution, the fluctuating ionizing field makes less of a difference to the required $\langle \Gamma \rangle$.  Remember, however, that these relatively small changes are always swamped by the differing $\kappa$'s, and a fluctuating background always requires a larger $\langle \Gamma \rangle$ than the uniform case.

For $z < 2.7$, the normalized points lie near the HRH07 curve. The averaged opacity at $z = 2.25$, which is significantly lower, relies on a single line of sight, HE 2347-4342, and differs significantly from the trend seen in Figure~\ref{fig:expt}. As expected, the inferred $\Gamma$ fluctuates considerably over this redshift range.  In part, these variations are due to the limited amount of data, both in the number and redshift coverage of usable quasar sightlines.   But the UV background fluctuates considerably, especially during reionization (see the $f(\Gamma)$ discussion in \S\ref{sec:semi}). For $z > 2.8$, the calculated photoionization rate consistently undershoots the model prediction, possibly indicating the end of helium reionization around that time: there is much more \ion{He}{2} than can be accommodated by a smoothly varying emissivity or attenuation length.

\subsection{ Evolution of the attenuation length } \label{sec:R}

\begin{figure}
   \plotone{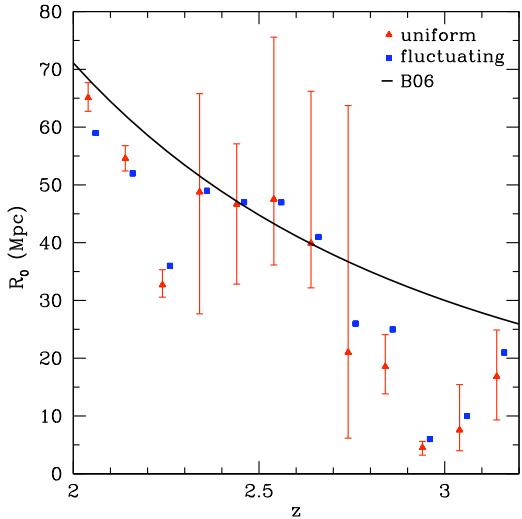}
   \caption{ Evolution of the helium-ionizing attenuation length with redshift. The calculated mean free path generally increases with a discontinuity around $z = 2.8$. The fluctuating (squares) background appears to smooth the evolution as compared to the uniform (triangles) background. The results are matched to the \citet{Bolt06} attenuation length (solid curve) at $z = 2.45$. The estimated uncertainties are shown only for the uniform case but are similar for both. }
   \label{fig:R_0}
\end{figure}

Because the measured quasar emissivity evolves smoothly with redshift, the most natural interpretation of this discontinuity is in terms of the attenuation length, which intuitively evolves rapidly at the end of reionization when \ion{He}{3} regions merge together and sharply increase the horizon to which ionizing sources are visible. Following the prescription for the UV background in \S\ref{sec:UVbg}, we calculate the mean free path $R_0$, given the HRH07 QLF and the $\Gamma_{-14}$ from the previous section. This procedure amounts to varying the solid curve, via $R_0$, to match the points in Figure~\ref{fig:Gamma}. The redshift evolution of the attenuation length for uniform (triangles) and fluctuating (squares) radiation backgrounds is plotted in Figure~\ref{fig:R_0}, with equation~(\ref{eq:Ro}) as a reference. The normalization remains the same as the previous section, i.e. $z = 2.45$ is the fiducial point.

Similarly to the inferred photoionization rate, the points vary about the reference curve for $z < 2.7$ and depart from it for $z > 2.8$. The uncertainties, which are shown only for the uniform UV background (but are comparable for the other case), are again quite large. Incorporating fluctuations reduces the severity of, but does not eliminate, the jump in the evolution of the attenuation length. Once again the results lie consistently below the curve for higher redshifts. From the viewpoint of $\Gamma$ or $R_0$, there appears to be a systematic change in behavior above $z \approx 2.8$. The marked decrease in $R_0$ that is required, by at least a factor of two from the fiducial model, indicates an important change in the state of the IGM.  However, as we have described above, a single attenuation length is no longer appropriate during reionization, so in \S\ref{sec:toy} we will turn to models of inhomogeneous reionization.

\subsection{The IGM Temperature-Density Relation} \label{sec:T}

As discussed in \S\ref{sec:semi}, the temperature-density relation of the IGM is a complicated question and an important component of the semi-analytic model. For the majority of the paper, we assume an isothermal model, i.e. $\gamma = 1$ in $T = T_0\Delta^{\gamma - 1}$. In reality, the temperature may depend on the density of the IGM. A further complication arises during (and shortly after) helium reionization when the IGM is inhomogeneously reheated and subsequently relaxes to a power law \citep{Gles05, Furl08a, McQu09}. 

\begin{figure}
   \plotone{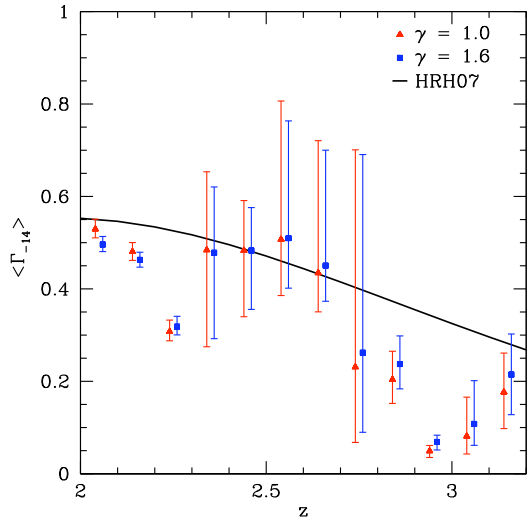}
   \caption{Comparison of the inferred photoionization rate (in units of $10^{-14}$~s$^{-1}$) for two temperature-density relations, given a homogeneous radiation background. The triangular points are derived using an isothermal model. The square points assume a steeper temperature-density relation, $\gamma = 1.6$. The points have a small redshift offset for illustrative purposes. The photoionization rate computed from the HRH07 quasar luminosity function is plotted for reference. Both models are normalized to the observations at $z=2.45$.   }
   \label{fig:T}
\end{figure}

To partially address the former issue, we repeat the calculation of $\Gamma$ for a homogeneous radiation background, but now with $\gamma = 1.6$, shown in Figure~\ref{fig:T}. The difference between the two cases is $\lesssim 30\%$. Overall, the steeper temperature-density relation slightly smoothes the jump in $\Gamma_{-14}$ near redshift $z \approx 2.8$ but is insufficient to fully explain the observed disconitnuity. As expected, the normalization differs between the two equations of state: $\kappa_{\gamma = 1.0} = 0.457$ and $\kappa_{\gamma = 1.6} = 0.277$ (see Fig.~\ref{fig:kappa}). In other words, a model with a higher $\gamma$ requires a higher $\langle \Gamma \rangle$ to achieve the same optical depth. This is because a steeper temperature-density relation makes the low-density IGM, which dominates the transmission, colder and hence \emph{more} neutral. Overall, then, the temperature-density relation and fluctuating ionizing background lead to a systematic uncertainty of nearly a factor of four in the mean photoionization rate inferred from the \ion{He}{2} forest.

\section{ Models for \ion{He}{2} reionization } \label{sec:toy}

We have seen that the observations appear to require a genuine discontinuity in the properties of the IGM at $z \approx 2.8$, although large statistical errors stemming from the small number of lines of sight prevent any strong conclusions. This change is often attributed to reionization; here we investigate this claim quantitatively with several ``toy" models for the evolution of the helium ionized fraction $\bar{x}_{\rm{HeIII}}$. This fraction determines $f(\Gamma)$ as described in \S\ref{sec:semi}. Assuming the HRH07 QLF and the attenuation length given by equation~(\ref{eq:Ro}),\footnote{Again, we note that during helium reionization the attenuation length should be evaluated with reference to individual quasars; in that case, it does not evolve strongly with redshift \citep{Furl08}.  This consideration will only strengthen our conclusions.} we calculate the effective optical depth via equation~(\ref{eq:F}). The fiducial point for normalizing $\kappa$ remains at $z = 2.45$. 

\begin{figure}
   \plotone{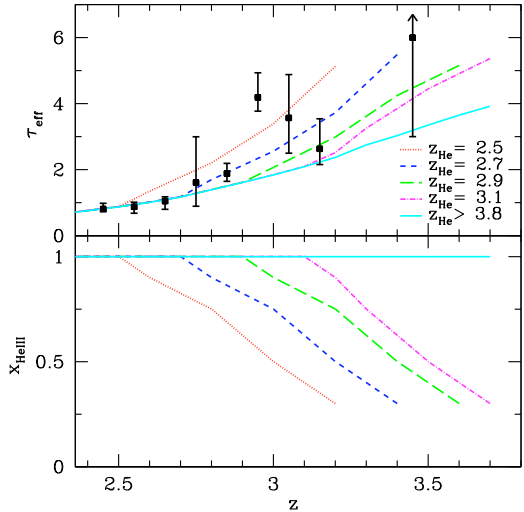}
   \caption{ The effective helium optical depth and \ion{He}{3} fraction for five toy reionization models. From left to right, the curves correspond to helium fully-ionized by $z_{\rm{He}} = 2.5, 2.7, 2.9, 3.1$ and post-reionization (or $z_{\rm{He}} > 3.8$). The \ion{He}{3} fraction evolution is varied slightly between models. The measured opacities are included for reference, including the lower limit at $z = 3.45$ from SDSS J2346-0016 \citep{Zhen04a}.}
   \label{fig:toys}
\end{figure}

Figure~\ref{fig:toys} shows the effective optical depth $\tau_{\rm{eff}}$ and ionized fraction $\bar{x}_{\rm{HeIII}}$ as a function of redshift for five reionization models. Each scenario is characterized by the redshift, $z_{\rm{He}}$, at which $\bar{x}_{\rm{HeIII}}$ reaches 1.0. From left to right in the figure, the curves correspond to $z_{\rm{He}} = 2.4, 2.5, 2.7, 3.1$ and $z_{\rm{He}} > 3.8$, i.e. post-reionization for the entire redshift range in question. The rate of ionization varies slightly between the models.\footnote{The unevenness in the curves arises because generating our Monte Carlo distributions is relatively expensive computationally, so we only generated a limited number at $\bar{x}_{\rm{HeIII}} = (0.3, 0.5, 0.75, 0.9,$ and 1.0) for each redshift.}  The measured opacities are plotted for reference, including the recently discovered SDSS J2346-0016 at $z=3.45$ \citep{Zhen04a,Zhen08}. Note that we do not estimate any cosmic variance uncertainty. 

The effective optical depth evolves smoothly in the post-reionization regime, which seems compatible with the data below $z \approx 2.8$. The $z_{\rm{He}} = 2.7$ model seems to fit the data best, but a range of values for the reionization redshift would be consistent with the existing data. Note that models with $z_{\rm He} > 3$ seem to evolve too smoothly; however, none of the curves displays enough of a discontinuity to match the data completely.  Obviously, more lines of sight at $z \ga 3$ are needed to reduce the wide cosmic variance.

\section{ Discussion } \label{sec:disc}

We have applied a semi-analyic model to the interpretation of the \ion{He}{2} \Lya forest, one of the few direct observational probes of the epoch of helium reionization. Using simple assumptions about the IGM, the ionization background, and our empirical knowledge of quasars, we have inferred the evolution of the helium phoionization rate and the attenuation length from the \ion{He}{2} effective optical depth. We averaged the opacity measurements over 5 sightlines, which show an overall decrease in $\tau_{\rm eff}$ with decreasing redshift and a sharp jump at $z \approx 2.8$ with the alternating low and high absorption at higher reshifts. After proper normalization, our model provides good agreement to the lower redshift data, but -- assuming smooth evolution in the quasar emissivity and attenuation length -- consistently overpredicts $\langle \Gamma \rangle$ above $z \approx 2.8$. Although the uncertainties are large, these results suggest a rapid change in the IGM around that time.

Our semi-analytic model is based the quasar luminosity function and the helium-ionizing photon attenuation length, which are determined empirically. The uncertainty in these quantities significantly affects our results. In particular, the plausible range of the mean EUV spectral index, $1.6 \lesssim \alpha \lesssim 1.8$, shifts the amplitude of our results by a factor of about two. Our model takes only the mean value and does not account for the variation in $\alpha$ from different quasars. Furthermore, our treatment of the attenuation length ignores any frequency dependence. However, these factors likely only affect the amplitude, not the evolution, of the inferred photoionization rate, especially the the jump at $z \approx 2.8$. A steeper redshift evolution for the attenuation length would decrease the severity of the feature around $z \approx 2.8$, but a simple power law cannot eliminate it, given our method.

In calculating $\Gamma$ and $R_0$, we compared uniform and fluctuating backgrounds. Although helium reionization is thought to be inhomogeneous (see \citealt{Furl08}), the assumption of a uniform background has been common. We found that the uniform case produces an effective optical depth approximately a factor of two smaller for a fixed $\langle \Gamma \rangle$. Thus, properly incorporating the fluctuating background is crucial for interpreting the \ion{He}{2} forest in terms of the ionizing sources. Furthermore, we find that the inclusion of background variations slightly smoothes, but does not remove, the jump in the attenuation length at $z \approx 2.8$. A clear change in the IGM does appear to occur around this redshift.

The discontinuous behavior in $\Gamma$ and $R_0$ led us to include helium reionization in our model through the distribution $f(\Gamma)$. During reionization, the ionized helium fraction determines this distribution, and we studied several toy models for the redshift evolution of $\bar{x}_{\rm{HeIII}}$. These models suggest $z_{\rm{He}} \approx 2.7$ as the best fit to the data, but the statistical uncertainties in $\tau_{\rm{eff}}$ are large. We do not account for cosmic variance and only consider the mean effective optical depth. Our method also makes assumptions that are not valid during reionization, e.g. a power-law temperature-density relation, but this does not appear to affect the discontinuity significantly.

In fact, the most important caveat to our model is the use of the fluctuating Gunn-Peterson approximation, which is a simplified treatment of the \Lya absorption.  The approach ignores the wings of absorption lines, peculiar velocities, and line blending; overall, these effects require us to add an unknown renormalization factor (of order $\sim 0.3$--$0.5$) when translating from $\Gamma$ to optical depth and compromises attempts to measure the absolute value of $\langle \Gamma \rangle$.  One danger is the possible redshift evolution of this factor:  we have assumed that it does not evolve, but in reality the line structure and temperature of the forest do evolve, especially at the end of reionization.  More detailed numerical simulations that incorporate both the baryonic physics of the \Lya forest and the large-scale inhomogeneities of helium reionization are required to explore this fully.

Interestingly, if our interpretation is correct then it appears that helium reionization completes at $z_{\rm He} \la 3$.  This places it several hundred million years \emph{after} the epoch suggested by indirect probes of the \ion{H}{1} \Lya forest.  Specifically, some measurements of the temperature evolution of the forest show a sharp jump at $z \approx 3.2$ and a shift toward isothermality \citep{Scha00, Rico00}, which has been interpreted as evidence for helium reionization \citep{Furl08a, McQu09}.  It is not clear whether this time lag can be made consistent; it probably depends on the details of the line selection in the observations (see the discussion in \citealt{Furl08a}).  Moreover, late helium reionization would present further difficulties for an explanation of the $z \approx 3.2$ feature in the \ion{H}{1} \Lya forest opacity in terms of helium reionization (see also \citealt{Bolt09}). Another indirect constraint is consistent, however, with our picture: reconstruction of the ionizing background from optically thin metal systems finds an effective optical depth in \ion{He}{2} \Lya photons slightly higher than the direct measurements, but with a similar redshift evolution \citep{Agaf05, Agaf07}.

The most significant limitation in the data is the relatively small number of lines of sight, producing large variations in the measured transmission, especially at $z \ga 3$ where the cosmic variance is large.  Fortunately, a number of new lines of sight have been found \citep{Zhen08,Syph09}, and the recent installation of the Cosmic Origins Spectrograph (COS) on the \emph{Hubble Space Telescope} adds a new instrument to our arsenal.  Although the nominal wavelength range of COS limits it to $z \ga 2.8$, this is precisely the most interesting range for studying reionization.  Our models show that $\tau_{\rm eff} \la 5$ so long as $x_{\rm HeIII} \ga 0.3$ (see Fig.~\ref{fig:toys}), so there should be a relatively wide redshift range with measurable transmission -- especially when considering the wide variations in the ionizing background expected during and after helium reionization (see Fig.~\ref{fig:f(j)} and \citealt{Furl08b}).

Finally, these prospects point out one important difference between \ion{He}{3} and \ion{H}{1} reionization:  the near-uniformity of the ionizing background at the end of \ion{H}{1} reionization means that very little residual transmission can be expected at $z \ga 6$ for that event, making the \Lya forest relatively useless for studying reionization.  In contrast, the large variance intrinsic to the \ion{He}{2}-ionizing background produces much stronger fluctuations and makes the epoch of reionization itself accessible with the \ion{He}{2} \Lya forest.

Another interesting difference between helium and hydrogen is the effect of including fluctuations on the photoionization rate inferred from the \Lya forest. We find that assuming a uniform ionizing background \emph{underestimates} $\Gamma$ by up to a factor of two, while during hydrogen reionization the effect is much smaller -- only a few percent \citep{Bolt07, Mesi09}. During and after helium reionization, the fluctuations are much more pronounced than the hydrogen equivalent, leading to a much broader $f(J)$, so that more of the Universe lies significantly below the mean. For hydrogen reionization, the distributions are much narrower, favoring $\Gamma$ near the mean.  In addition, after hydrogen reionization the density distribution is much wider than $f(\Gamma)$, so that the latter provides only a small perturbation; the opposite is true in our case.

Our general approach is very similar to \citet{Fan02} and \citet{Fan06}, who also interpreted the \ion{H}{1} \Lya forest data at $z \sim 6$ using a uniform ionizing background and the same IGM density model as we have (although in their case that model required extrapolation to the relevant redshifts).  They also found a discontinuity in the optical depth (at $z \sim 6.1$), which is often taken as evidence for \ion{H}{1} reionization.  But during this earlier epoch, that inference is less clear because of the near saturation of the forest and the unknown attenuation length (whose evolution really determines the overall ionizing background, but which may evolve rapidly even after reionization; \citealt{Furl09}).  Nevertheless, we hope that understanding this discontinuity in the \ion{He}{2} forest properties will shed light on the problem of hydrogen reionization.

We thank J.~S.~Bolton, J.~M.~Shull, and J.~Tumlinson for sharing their data in electronic form.  This research was partially supported by the NSF through grant AST-0607470 and by the David and Lucile Packard Foundation.

\bibliographystyle{apj}
\bibliography{helium_reionization}

\end{document}